\documentclass[%
 reprint,
 amsmath,amssymb,
 aps,
]{revtex4-2}

\usepackage{graphicx}% Include figure files
\usepackage{dcolumn}% Align table columns on decimal point
\usepackage{bm}% bold math
\usepackage{subfigure}
\begin{document}

\preprint{APS/123-QED}

\title{Measurement of the Per Cavity Energy Recovery Efficiency in the Single Turn CBETA Configuration}% Force line breaks with \\
%\thanks{A footnote to the article title}%

\author{C. Gulliford}
\author{N. Banerjee}
%\author{J. Barley}
\author{A. Bartnik}
%\author{I. Bazarov}
%\author{D. Burke}
\author{J. Crittenden}
%\author{L. Cultrera} 
\author{K. Deitrick}
\author{J. Dobbins}
%\author{C. Franck}
%\author{R. Gallagher}
%\author{B. Heltsley}
\author{G. H. Hoffstaetter}
%\author{D. Jusic}
%\author{R. Kaplan}
%\author{V. Kostroun}
%\author{Y. Li}
%\author{M. Liepe}
%\author{W. Lou}
\author{P. Quigley}
%\author{D. Sabol}
%\author{D. Sagan}
%\author{J. Sears}
%\author{C. Shore}
%\author{E. Smith}
\author{K. Smolenski}
%\author{V. Vesherevich}
%\author{D. Widger}
\affiliation{Cornell Laboratory for Accelerator Based Sciences and Education,
\\ 
Cornell University,
\\
Ithaca, New York 14850, USA }

\author{J.{~}S. Berg}
%\author{S. Brooks}
%\author{R. Hulsart}
\author{R. Michnoff}
\author{S. Peggs}
\author{D. Trbojevic}
\affiliation{Brookhaven National Laboratory,\\
Upton, NY 11973-5000, USA}
%\author{C. Gulliford}

%\collaboration{Cornell University}%\noaffiliation

%\author{BNL people here}
% \homepage{http://www.Second.institution.edu/~Charlie.Autho%r}
%\%affiliation{
% Second institution and/or address\\
% This line break forced% with \\
%}%
%\affiliation{
% Third institution, the second for Charlie Author
%}%
%\author{BNL Authors here}
%\affiliation{%
% Authors' institution and/or address\\
% This line break forced with \textbackslash\textbackslash
%}%

%\collaboration{CLEO Collaboration}%\noaffiliation

\date{\today}% It is always \today, today,
             %  but any date may be explicitly specified

\begin{abstract}
Prior to establishing operation of the world's first mulit-turn superconducting Energy Recovery Linac, %(ERL)
 the Cornell-BNL Energy Recovery Test Accelerator %(CBETA)
was configured for one turn energy recovery.  In this setup, direct measurement of the beam loading in each of the main linac cavities demonstrated high energy recovery efficiency.  Specifically, a total one-turn power balance efficiency of 99.4\%, with per cavity power balances ranging from 99.2-99.8\%, was measured. When accounting for small particle losses occurring in the path length adjustment sections of the return loop, this corresponds to per cavity single particle energy recovery efficiencies ranging from 99.8 to 100.5\%.  A maximum current of 70 $\mu$A was energy recovered, limited by radiation shielding of the beam stop in its preliminary installation.
\end{abstract}

%\keywords{Suggested keywords}%Use showkeys class option if keyword
                              %display desired
\maketitle

%\tableofcontents

\section{\label{sec1:level1}Introduction}

CBETA, the Cornell-BNL Energy recovery linac Test Accelerator, the world's first superconducting radiofrequency (SRF) multi-turn energy recovery linac (ERL)~\cite{ref:4pCBETA}, was designed to study phenomena important to the ERL community, including generation and energy recovery of high current beams, %with currents up to 100 mA (single-turn configuration) or 40 mA (4-turn configuration), 
the Beam Breakup (BBU) instability \cite{ref:bbu:lou}, halo development and collimation, as well as growth in energy spread by Coherent Synchrotron Radiation (CSR) and microbunching \cite{ref:csr:lou}. Understanding of ERLs is of particular relevance for an electron-ion collider (EIC), which is currently one of the highest priorities for the nuclear physics and accelerator communities~\cite{ref:LongRange1,ref:NAS}: specifically, meeting the nuclear physics requirements for an EIC requires ion cooling using high brightness, high current electrons with micron normalized emittances and with roughly 100 mA beam currents; ERL technology is the best candidate for producing the required electron beam.

This work reports on the establishment of high efficiency energy recovery (ER) in the CBETA single-turn configuration. First, the experimental set-up is described, followed by a brief discussion of preliminary measurements undertaken to calibrate and characterize the machine before establishing ER, including how the linac cavity voltage is calibrated and the on-crest phases determined.  After this, the various figures of merit for ER efficiency are defined and discussed along with methods used to determine linac settings that yield high-efficiency single turn ER.  Following this, direct measurements of the beam loading in each of the linac cavities are analyzed, providing a per cavity measure of the ER efficiency. Finally, a brief discussion of raising the current in the single ER state is given.

\section{\label{sec2:level1}Experimental Setup}

\begin{figure}[!htb]
\centering
\includegraphics*[width=\columnwidth]{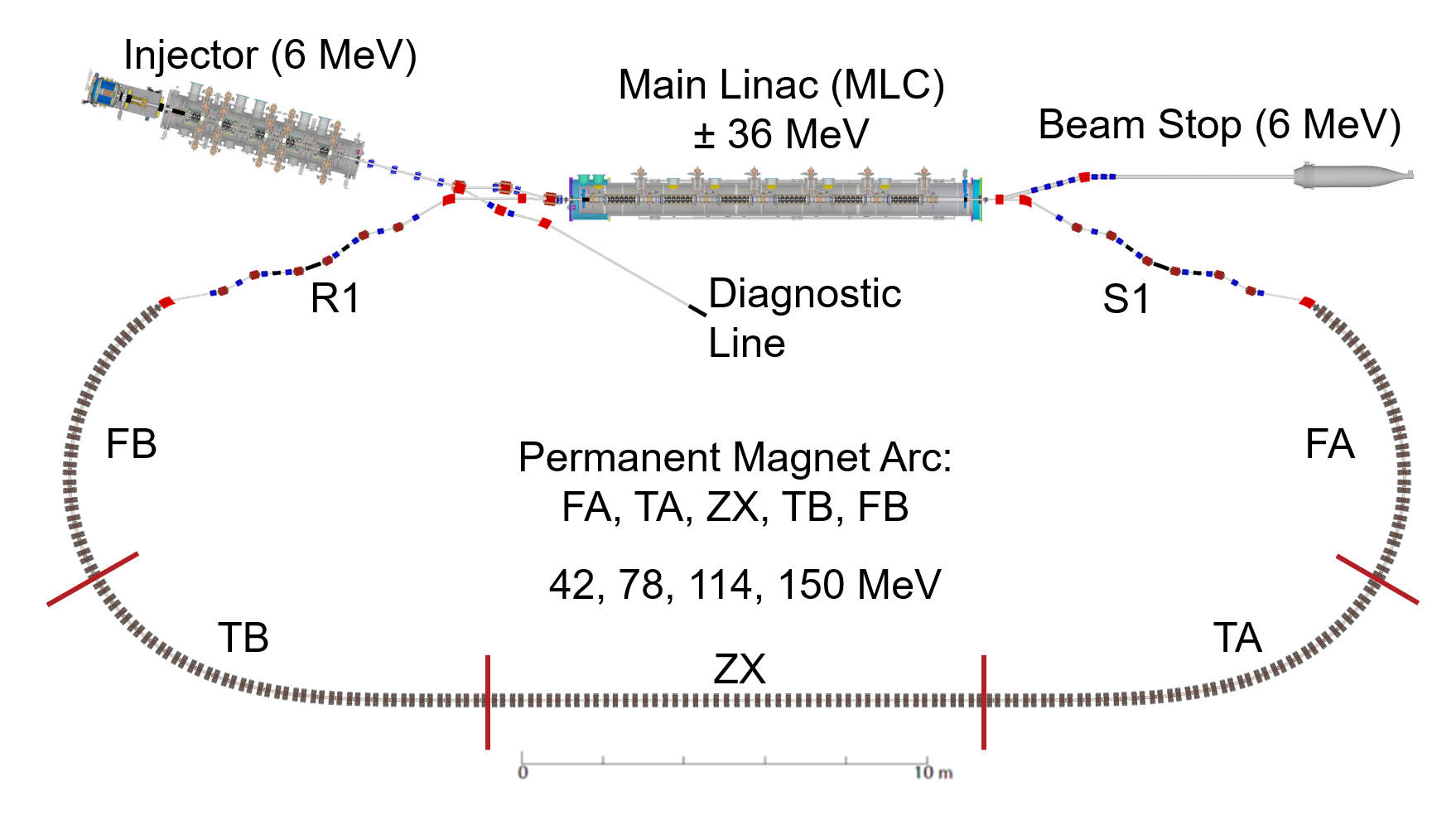}
\caption{Layout of CBETA in the one-turn configuration.}
\label{FIG:CBETA_LAYOUT_01}
\end{figure}

Fig.~\ref{FIG:CBETA_LAYOUT_01} shows the single turn configuration of CBETA.  In this configuration the Cornell injector delivers a 6 MeV beam \cite{ref:lowemitter, ref:lowemitter2,ref:lowemitter3, ref:hcrecord} either to the diagnostic line for 6D phase space characterization of the beam after the merge or to the main linac cyromodule (MLC) which nominally provides 36 MeV.  A splitter section (S1) follows the MLC and provides orbit, optics matching and path length adjustment for injection into the FFA return loop (FA, TA, ZX, TB, FB).  After the FFA return loop a recombination section (R1) provides further orbit, optics, and path length adjustment before the beam is decelerated in the MLC and sent to the beam dump section.

The CBETA FFA return loop transports a wide range of beam energies in a single beam line without the need to vary the magnetic fields using a Non-scaling linear gradient (FFA-LG) design \cite{ref:Yoshimoto,ref:Mashida1,ref:brooks,ref:mashida}, and features other novel technologies, such as the use of Halbach combined-function permanent magnets~\cite{Brooks:2017fvi, Brooks:2019rgg} and an adiabatic transition between the arc and straight sections~\cite{berg:erl17}.

\section{\label{sec3:level1}Measurements and Results}

The establishment of energy recovery requires precise control of the phasing and voltage calibration of the linac cavities, as well as the orbit and optics functions around the loop.  The following sections describe the measurement techniques for determining the linac voltage calibrations and on-crest phases as well as verification of the orbit correction and linear optics through the FFA return loop.  Verification of the linear optics in the conventional splitter magnets in the S1/R1 sections was performed using standard response techniques \cite{ref:FAT}, and is not included here for brevity.

\subsection{Preliminary}

\subsubsection{Main Linac Voltage and Phase Calibration}

%\begin{figure}[!htb]
%\centering
%\includegraphics*[width=0.85\columnwidth]{phase6-crop}
%\caption{Layout of CBETA in the one-turn configuration.}
%\label{FIG:CBETA_LAYOUT_01}
%\end{figure}
\begin{figure}[ht]
    \begin{center}
        \subfigure[\hspace{0.2cm}]{%
           \label{fig:phase6}
           \includegraphics[width=0.35\textwidth]{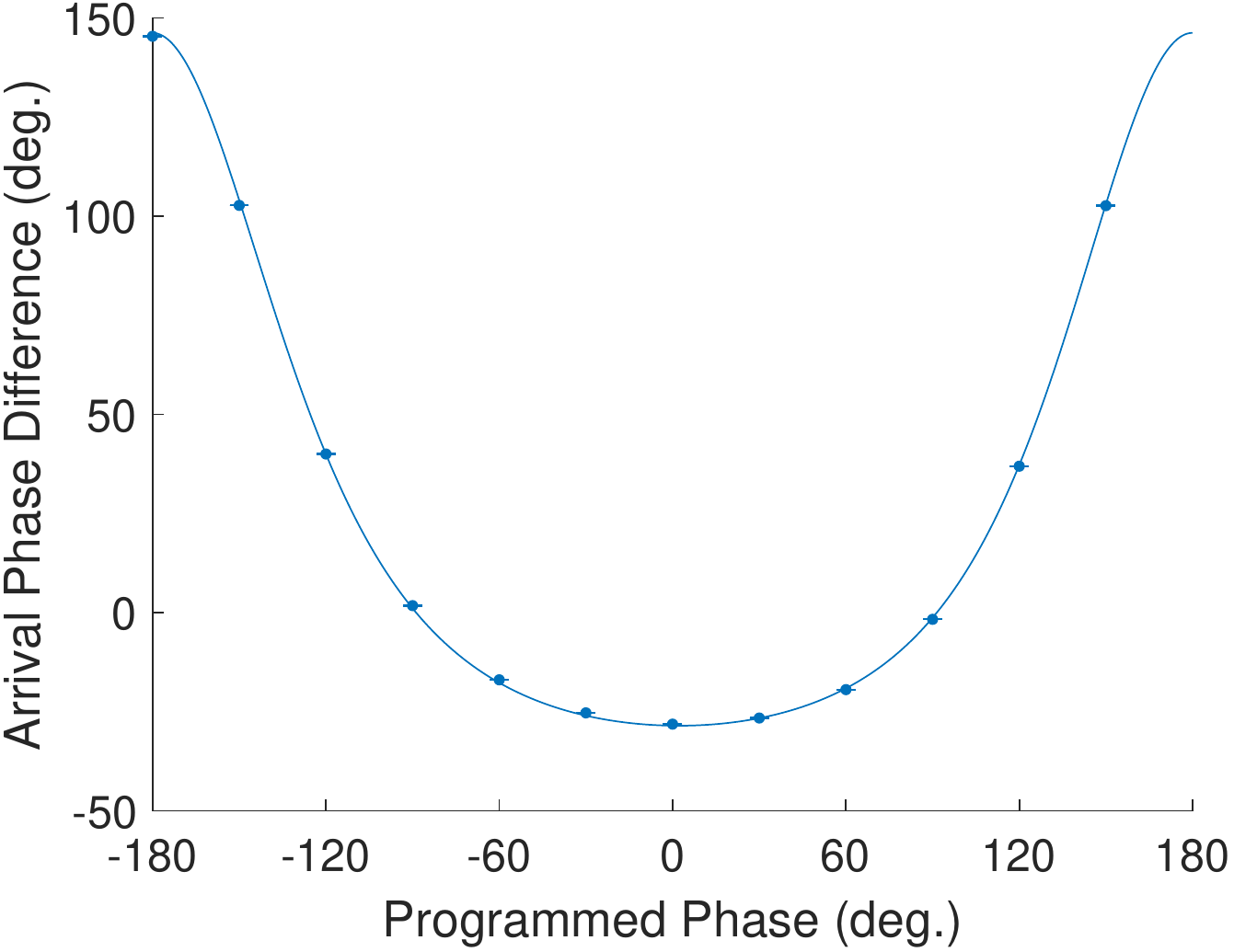}
        } %  ------- End of the first row ----------------------%
        \subfigure[\hspace{0.2cm}]{%
           \label{fig:phase1}
           \includegraphics[width=0.35\textwidth]{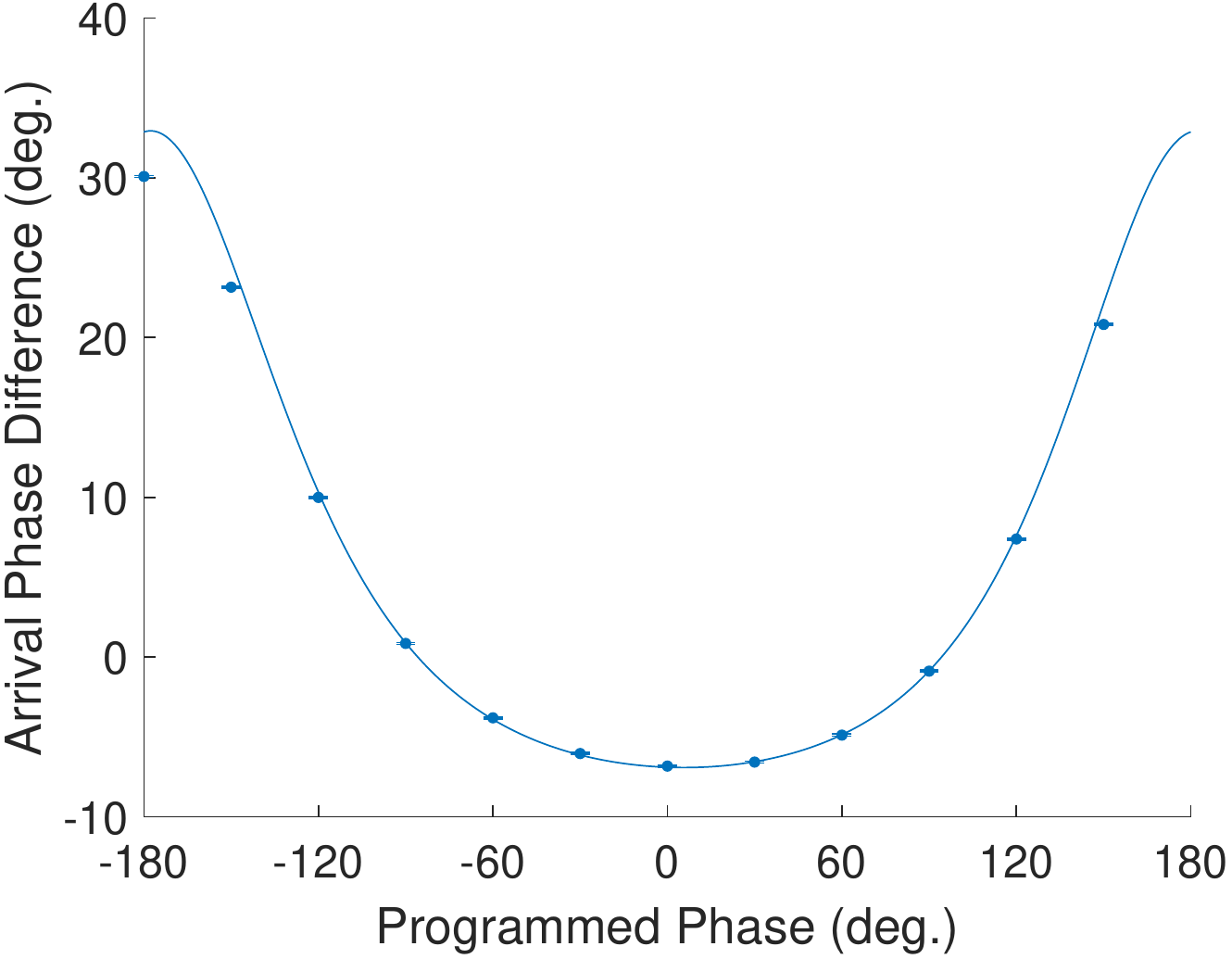}
        } %  ------- End of the first row ----------------------%
    \end{center}
    \caption{%
    \label{fig:phase61}
    BPM arrival phase difference measurements used for calibrating the phase offset for the first (a) and last linac cavity (b).  Solid line shows the best fit of the data to the MLC phasing model.}%
\end{figure}

Calibration of both the injector and MLC cavity phases was performed upon each machine start up. The calibration of each MLC cavity is performed by turning on only the cavity in question, with the voltage set to approximately half of the incoming beam energy, scanning its phase through 360 degrees, and measuring the arrival time at a BPM just downstream of the MLC. After performing this measurement for each of the cavities, we compare the arrival time as a function of cavity phase to one computed from a model and use that to determine the difference between the programmed cavity phase and its actual phase, and adjust the phase calibration appropriately. The computation also determines the incoming beam energy. If desired, a voltage calibration factor can be determined for each cavity as well. Further details of the computation and measurement follow.

The model of the MLC tracks an on-axis particle from the BPM just before the linac to a BPM just after the linac. In principle the upstream BPM is not necessary, but simultaneous measurement of the upstream and downstream arrival times removes some noise from the measurements. The longitudinal positions of the cavities and the BPMs are known from survey.

Within cavity $k$, the energy $E$ and time $t$ evolve according to
\begin{align}
  \dfrac{dt}{ds} &= \dfrac{E}{c\sqrt{E^2-(mc^2)^2}}\\
  \dfrac{dE}{ds} &= qV_k\mathcal{E}(s)\cos(\omega t+\phi_k+\psi_k).
\end{align}
Here $s$ is the position along the cavity axis, $q$ is the particle charge, $m$ is its mass, $c$ is the velocity of light, and $\omega$ is the angular RF frequency of the cavity. $\mathcal{E}(s)$ is the the electric field along the cavity axis when the electric field is at its maximum, multiplied by $e$, and divided by the maximum energy gain the cavity would give a particle with $q=e$ moving with velocity $c$ and has units of inverse length. $\mathcal{E}(s)$ is computed from a finite-element computation. $V_k$ and $\phi_k$ are the voltage and phase, respectively, that the control system sets for the cavity; our goal is to calibrate these quantities. $\psi_k$ is a phase offset related to our definition of zero phase. $\psi_k$ is defined so that when a particle comes in with an energy of $E_{\text{ref}}=6\text{ MeV}$, the energy gain is maximized when $\phi_k=0$ in the limit of $V_k\to0$. $\psi_k$ depends on the position of the cavity relative to the BPMs and thus is subscripted with the cavity index.

Arrival times are measured at the BPMs as phases relative to the 1.3~GHz RF frequency and are denoted as $\theta$. For a given cavity $k$ and a given cavity phase and voltage setting $j$, the measurements are $\theta_{jk}^{\text{U}}$ at the upstream BPM and $\theta_{jk}^{\text{D}}$ at the downstream BPM. The arrival times with all cavities off are also measured and denoted $\theta_0^{\text{U}}$ and  $\theta_0^{\text{D}}$. The model gives a prediction of the time between the two BPMs as a function of voltage $V$ and phase $\phi$ at cavity $k$, which we call $T_k(V,\phi,E_0)$, where $E_0$ is the incoming energy. We assume there is an offset $\delta_k$ between the actual cavity phase and the phase requested from the control system, and a multiplicative factor $\lambda_k$ between the actual cavity voltage and the voltage given to the control system. We thus minimize
\begin{multline}
  \sum_{jk}
  \{\theta_{jk}^{\text{D}}-\theta_{jk}^{\text{U}}-\theta_0^{\text{D}}+\theta_0^{\text{U}}\\ - \omega[T_k(\lambda_k V_{jk},\phi_{jk}+\delta_k,E_0)-T_k(0,0,E_0)]\}^2
\end{multline}
%with respect to $\lambda_k$, $\delta_k$, and $E_0$, 
where $V_{jk}$ and $\phi_{jk}$ are phases and voltages given to the control system for the measurements with cavity $k$ on. This minimization can be computed very rapidly because a good initial guess can be obtained from a model where the cavity is approximated as a single energy kick with some additional simplifications, and derivatives of the model with respect to parameters can be computed by integrating the equations of motion for the derivatives of $T_k$ with respect to those parameters.

Each time the machine is turned on, this measurement was performed in order to determine the phase offsets $\delta_k$. For this computation, we leave out the $\lambda_k$ but do find the incoming energy $E_0$. The $\delta_k$ can be found using this method with an accuracy ranging from a couple tenths of a degree (cavities further upstream) to about a degree (cavity furthest downstream) with a relatively small number of measurements (typically we scan in 30 degree steps with each cavity set to 3~MV for a 6~MeV incoming beam). Computing the $\lambda_k$ was only done a couple of times; it is best done with multiple (and larger) voltages, typically 3~MV and 4~MV, and requires a somewhat more involved process to update the cavity calibrations.

Figures~\ref{fig:phase6} and \ref{fig:phase1} show measurements of the arrival phase as a function of cavity phase, minus the arrival phase with the cavities off, for the first and last MLC cavities. These are typical of the calibration measurements that would be performed daily. The statistical errors in the measurements are negligible, the differences are dominated by systematic errors; the 
%most likely 
source of this error is alignment issues with the beam and cavities. Despite what appears to be a significant deviation of the model from the data for the cavity at the end of the MLC, more detailed analysis indicates that the error is only about 1~degree. In fact, detailed analysis of that last cavity indicates there is a voltage calibration error leading to some of the difference seen in Fig.~\ref{fig:phase1}. The phasing of the earlier cavities is more accurate than the later cavities due to the longer distance between the cavity and the downstream BPM.

%\subsubsection{1 Turn FFA Orbit Correction \label{oc}}

%Clean transmission through the FFA return loop requires correct injection onto the periodic orbit at the FFA entrance.  

\subsubsection{Orbit Correction and Tune Measurement\label{ssec:escan}}

Measurements to verify the orbit and linear optics of in the FFA return loop were performed as a final step before establishing energy recovering. Orbit correction was performed using an SVD based approach \cite{ref:CBETAVM}.  This algorithm uses the corrector to BPM response matrix served live in EPCIS from our online Bmad model of the machine and includes automatic finding of the periodic orbit in the arc sections of the return loop, and assumes the target orbit in the transition sections TA and TB is proportional to the model orbit. 
\begin{figure*}[ht]
    \begin{center}
        \subfigure[\hspace{0.2cm}Uncorrected orbit.]{%
           \label{fig:badorbit}
           \includegraphics[width=0.8\textwidth]{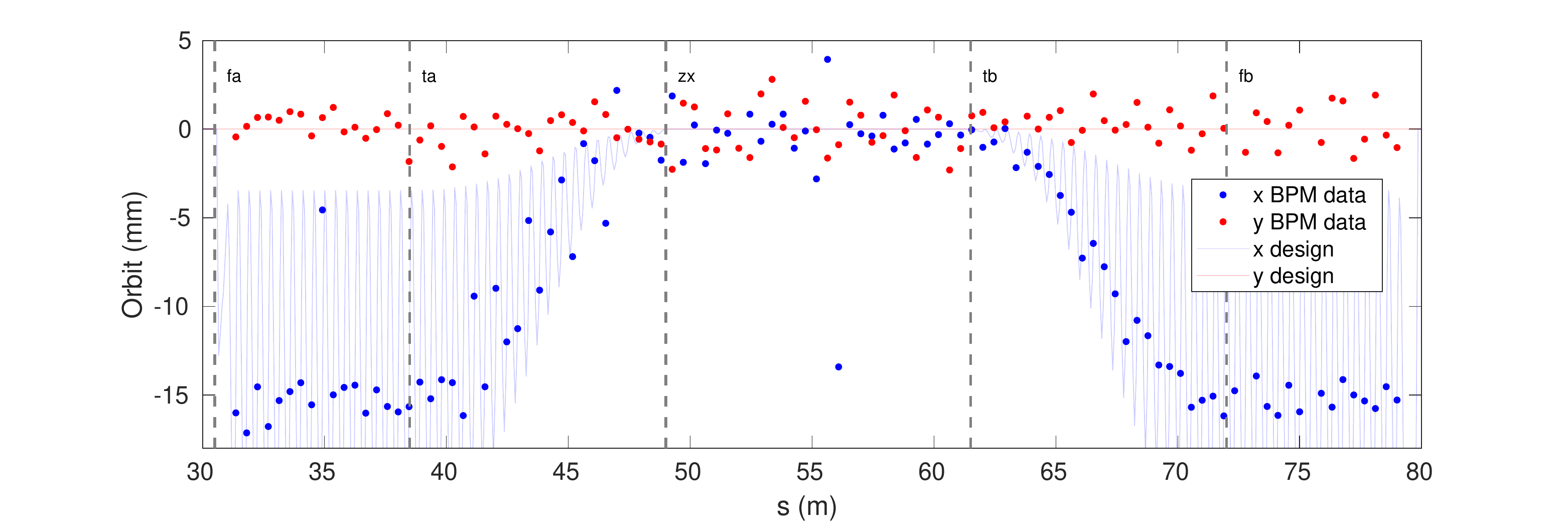}
        }\\ %  ------- End of the first row ----------------------%
        \subfigure[\hspace{0.2cm}Corrected orbit.]{%
           \label{fig:goodorbit}
           \includegraphics[width=0.8\textwidth]{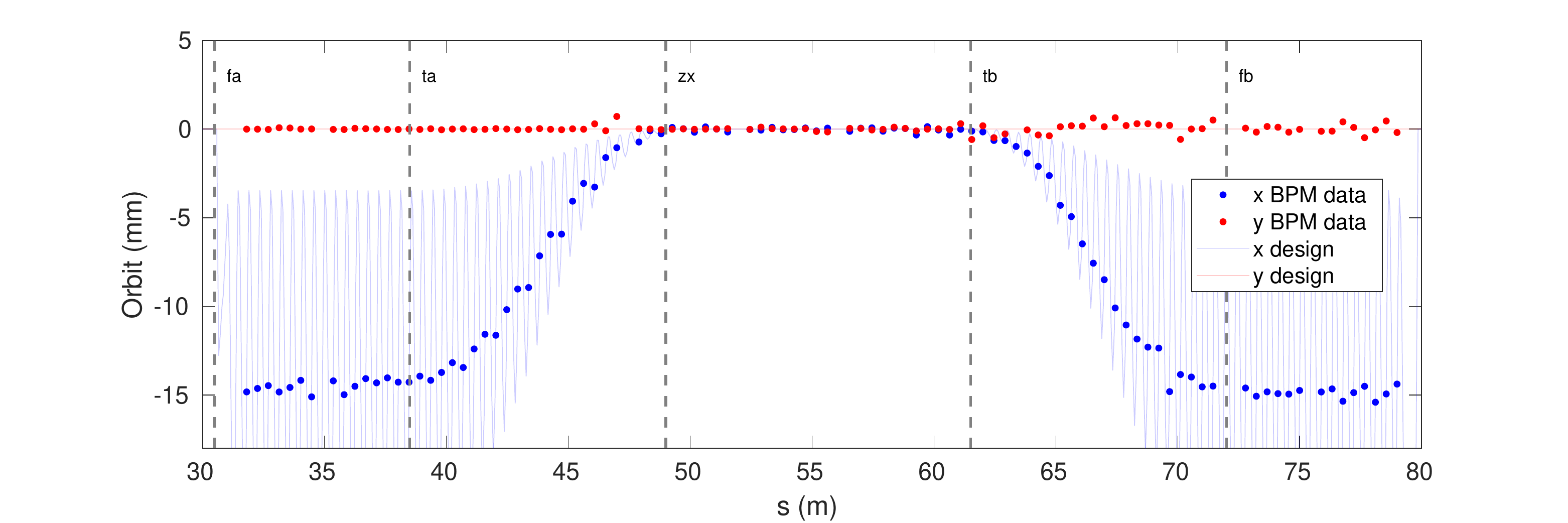}
        } %  ------- End of the first row ----------------------%
    \end{center}
    \caption{%
    \label{fig:orbit_correction}
    Online 1-pass orbit correction using SVD: (a) an example initial orbit tuned by hand (b) an orbit tuned using the SVD algorithm. The solid blue and red lines indicate the design orbit in the online model..}%
\end{figure*}

Fig.~\ref{fig:badorbit} shows a typical single pass orbit as measured on the FFA BPMs tuned by hand.  Here the horizontal and vertical orbit data on the BPMs is shown in blue and red respectively. Also shown is the theoretical orbit as computed from the online model. Applying the SVD algorithm section by section results in the orbit data shown in Fig.~\ref{fig:goodorbit}.  The fact that the algorithm would not converge when attempting correction globally through the entire machine suggests some discrepancy between the machine and modeled phase advance per cell (referred to as tunes). 

\begin{figure}[!hb]
    \begin{center}
        \subfigure[\hspace{0.2cm}FA/FB Arc Cell Phase Advances/$2\pi$]{%
           \label{fig:tunes_arc}
           \includegraphics[width=0.42\textwidth]{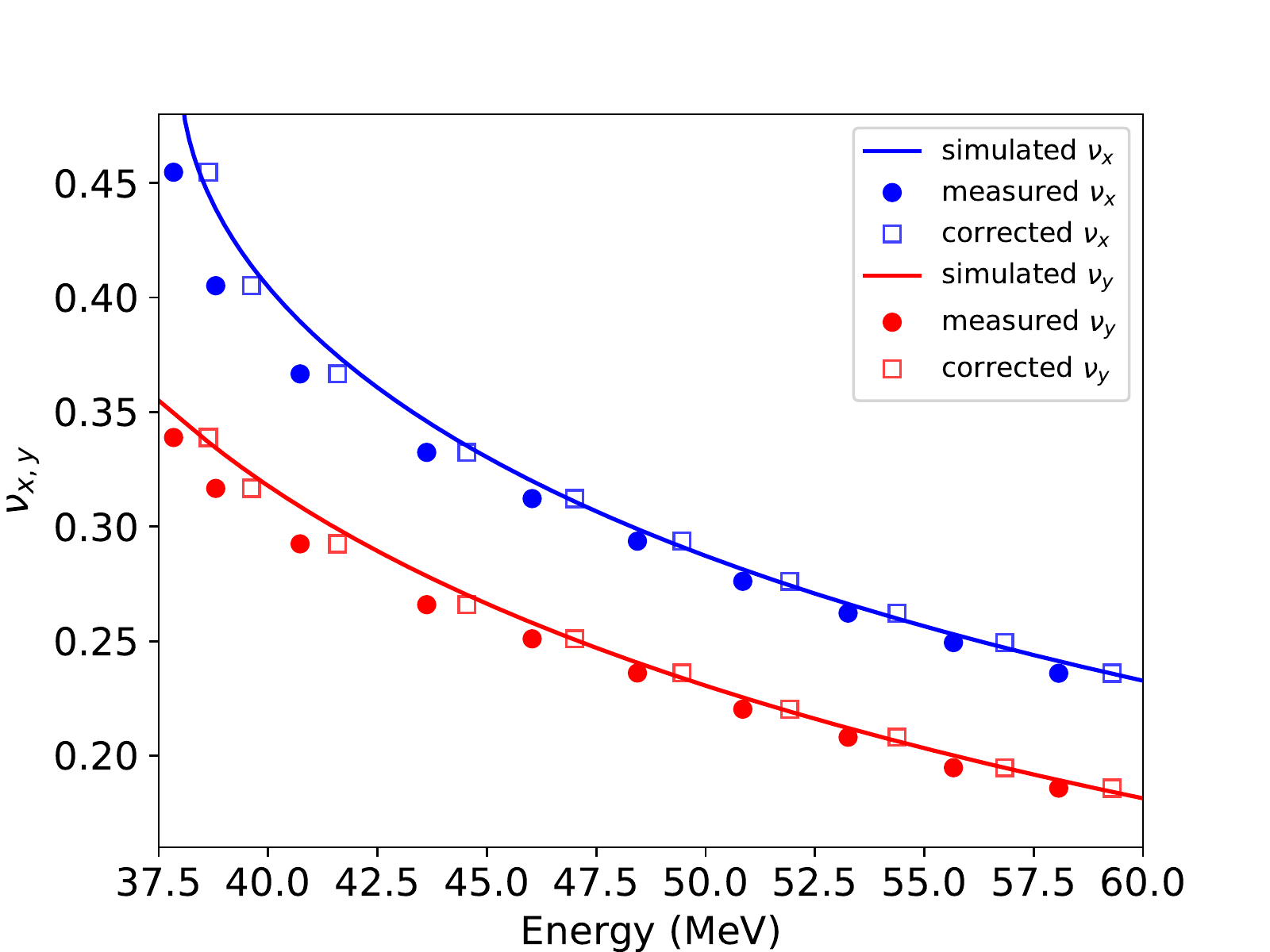}
        }\\ %  ------- End of the first row ----------------------%
        \subfigure[\hspace{0.2cm}ZX Cell Phase Advances/$2\pi$]{%
           \label{fig:tunes_zx}
           \includegraphics[width=0.42\textwidth]{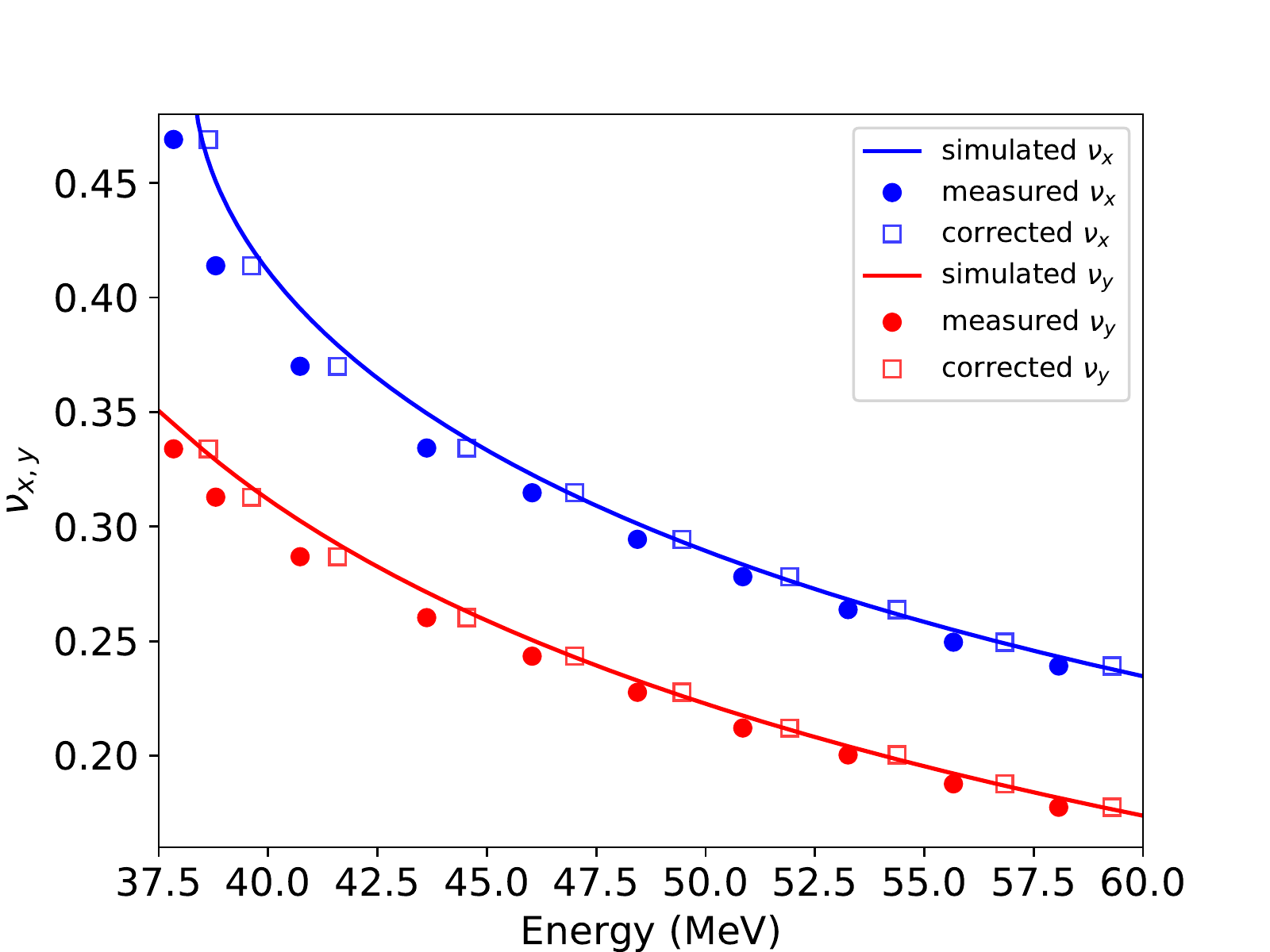}
        } %  ------- End of the first row ----------------------%
    \end{center}
    \caption{%
    \label{fig:tunes}
    Phase advance per cell per 2$\pi$ in the FA and FB arc sections (a) and in the ZX straight section (b). Measured data are shown with circular markers. Square markers indicate the measured data with the energy scaled by a factor of 1.02.}%
\end{figure}

This discrepancy between the model and measured tunes was further investigated by direct measurement of the tunes. This was accomplished by generating both horizontal and vertical orbit kicks in the S1 splitter which then propagated through the FFA.  The motion of the beam on the FFA BPMs was then analyzed using the method discussed in \cite{ref:FAT} resulting in phase advance through the cells in the FA, ZX, and FB sections of the FFA return loop.  This was performed for several different energies ranging from 39 to 59 MeV.  Fig.~\ref{fig:tunes} shows the comparison of the resulting horizontal phase advance/$2\pi$ per cell to the predicted values from particle tracking. The results for both arc sections (FA, FB), as well as the straight section (ZX) are shown in Fig.~\ref{fig:tunes_arc} and Fig.~\ref{fig:tunes_zx} respectively.  In all cases good agreement is seen with the corresponding simulated tunes. %with predictions from theory.  
%sThe agreement becomes better when fitting the data for an overall energy scale factor (here applied to the simulated curve).  

A similar technique for measuring the tunes has since been incorporated into the orbit correction algorithm as a way to generate a more accurate response matrix. These improvements allow for correction of the entire orbit in both the single or multi-turn CBETA configurations \cite{ref:4pCBETA}.

\subsection{Energy Recovery\label{ssec:EReff}}

The phases and voltages of the RF cavities during energy recovery operation are set with a number of goals in mind:
\begin{itemize}
\item The total energy at the end of the accelerating pass should be 42~MeV and the total energy at the end of the decelerating pass should be 6~MeV
\item In each individual cavity, the energy gained in the accelerating pass should be equal to the energy lost in the decelerating pass
\item The beam should be near the crest so as to avoid
  \begin{itemize}
  \item Sensitivity to beam timing and RF phase jitter
  \item Excessive increase in beam energy spread
  \end{itemize}
\end{itemize}
The first goal is redundant with the second, but is mentioned separately since in practice we will not meet the second goal but will nearly meet the first, for reasons discussed below. In this section we describe how we set the cavity phases for energy recovery operation.

The cavity phases and voltages used in the machine are determined from our model for the machine, which incorporates field maps for the cavities identical to those used for the calibration model described above. We describe two phases in this section: a phase that is set in both the model and the machine's control system (the ``set phase''), and an ``effective phase,'' which corresponds more closely to the intuitive notion of a cavity's phase.

Recall that each cavity's zero phase is calibrated using the 6~MeV injected beam. Zero phase is the phase for which a 6~MeV particle achieves its maximum energy gain for small cavity voltages. This calibration is done both in the real machine and the model. But for both the accelerating and decelerating pass, the energy of the beam going through the linac is higher than 6~MeV. For instance, if the beam were accelerated by 6 MeV in each of the first 5 cavities of the MLC, the set phase of the last cavity that would achieve the maximum energy gain in that cavity would be approximately $+37$~degrees.

To describe this difference between set phase and effective phase more precisely, we give a precise definition of effective phase of a cavity and a related quanitity we will call ``linac phase.'' First, define $E_{ij}$ to be the particle energy after pass $j$ through cavity $i$, with $E_{0j}$ being the energy before pass $j$ through the linac ($E_{0,j+1}=E_{6,j}$). These energies (other than $E_{01}$) are all a function the cavity voltages and phases. Similarly, define times $t_{ij}$ to be times at positions $s_i$ on pass $j$ through the linac such that the energy is $E_{ij}$ at time $t_{ij}$. The precise locations $s_i$ are unimportant, since we will only be taking derivatives with respect to $t_{ij}$. We then define the effective phase $\phi^{\text{cav}}_{ij}$ for cavity $i$ on pass $j$ so that the following equations hold
\begin{align}
  E_{ij}-E_{i-1,j} &= U^{\text{cav}}_{ij}\cos\phi^{\text{cav}}_{ij}\\
  \dfrac{dE_{ij}}{dt_{i-1,j}} &= -\omega U^{\text{cav}}_{ij}\sin\phi^{\text{cav}}_{ij}
\end{align}
The quantities $U^{\text{cav}}_{ij}$ are only used for the construction of $\phi^{\text{cav}}_{ij}$. The time derivatives are artificial in the sense that the arrival time at some intermediate point in the linac cannot be varied independently of arrival times earlier in the linac. Henceforth we will only describe these effective phases; the set phases for given effective phases are computed using the model described for the cavity phase calibration.

Similarly to the effective cavity phases, the linac phase $\phi^{\text{lin}}_j$ for pass $j$ is defined such that
\begin{align}
  E_{6j}-E_{0j} = U^{\text{lin}}_j\cos\phi^{\text{lin}}_j\\
  \dfrac{dE_{6j}}{dt_{0,j}} = -\omega U^{\text{lin}}_j\sin\phi^{\text{lin}}_j
\end{align}
The linac phase is in principle measurable, since the arrival time at the entrance to the linac can be varied by adjusting the sliding joints in the splitter lines, or for the first pass by changing the phases of all the RF cavities.

\begin{table}[htb]
\begin{center}
\caption{Effective cavity ($\phi^{\text{cav}}_{ij}$) phases and cavity voltages for theoretical perfectly energy balanced configuration.}
\begin{ruledtabular}
\begin{tabular}{l|rrrrrr}
    Cavity & 1 & 2 & 3 & 4 & 5 & 6\\\hline
    $\phi^{\text{cav}}_{i1}$ (deg.) &+3.97 & $-$4.06 & +0.06 & +0.99 & +4.54 & $-$5.49\\
    $\phi^{\text{cav}}_{i2}-\text{180}^\circ$ (deg.) & +5.47 & $-$4.51 & $-$0.98 & $-$0.06 & +4.09 & $-$3.99\\
    \hline
    $V_{c,i}$ (MV) & 6021 & 6013 & 6014 & 6014 & 6015 & 6022
  \end{tabular}
  \label{tab:phibal}
\end{ruledtabular}
\end{center}
\end{table}
\begin{table}[htb]
\begin{center}
\caption{Effective cavity ($\phi^{\text{cav}}_{ij}$) phases and cavity voltages for near energy balanced configuration used for ER measurements.}
\begin{ruledtabular}
\begin{tabular}{l|rrrrrr}
    Cavity & 1 & 2 & 3 & 4 & 5 & 6\\\hline
    $\phi^{\text{cav}}_{i1}$ (deg.) & -1 & -1 & -1 & -1 & -1 & -1\\
    $\phi^{\text{cav}}_{i2}-\text{180}^\circ$ (deg.) & +1.35 &  -0.53 & -1.08 & -1.03 & -0.32 & +1.99\\
    \hline
    $V_{c,i}$ (MV) & 6215 & 6205 & 6203 & 6202 & 6202 & 5001
  \end{tabular}
  \label{tab:phiunbal}
\end{ruledtabular}
\end{center}
\end{table}

Table~\ref{tab:phibal} shows the phases and voltages for the perfectly balanced configuration where energy gain in each cavity in the first pass equals the energy lost in the second pass and the overall linac phase is zero, so as to avoid inducing energy spread and to reduce the impact of any common timing jitter of the linac phases relative to injection. Unfortunately, this configuration created difficulties with beam stability in the return loop. This appeared to be related to RF phase stability and how far the beam was off-crest in some of the cavities: the further the beam is off-crest in a cavity, the more the energy changes in response to an RF phase change.

In practice this state was replaced by the one shown in Table~\ref{tab:phiunbal}. Here each cavity's voltage was set so that the energy gain during the acceleration pass through each cavity was 6.2~MeV, except for the last cavity which was set for 5~MV (the maximum achievable a the time of measurement).  The phases in each cavity were set to $-1$~deg.\ from the phase which would have given the maximum energy gain for the desired incoming energy. This very nearly corresponds to an effective cavity phase of $-1$ deg.\ in each cavity.  These settings result in the return beam being 1.8~keV above the 6 MeV injection energy as predicted by the model. %, which is lower than the measured energy resolution entering the dump line. 
With the beam in this nearly balanced ER state, we proceeded to quantify the individual cavity energy recovery efficiencies by direct measurement of the beam loading in each cavity. 

The general expression for the beam loading for a given cavity in terms of the forward power $P_+$ and reflected power $P_{-}$, the power dissipated in the cavity walls $P_c$, the beam current $I$, and the single particle voltage gain through the cavity is:
\begin{eqnarray}
P_b = P_+ - P_- - P_c = I\cdot\Delta V.
\label{eq:loading}
\end{eqnarray}
In the steady state and ultra-relativistic limit $P_c = V_c^2/2/(R/Q)Q_0$ and $\Delta V=V_c\cos\phi$ where $V_c$ is the on-crest ultra-relativistic cavity voltage, $Q_0$ is the intrinsic cavity quality factor, $R/Q$ is the ratio of the shunt impedance and quality factor, and $\phi$ is the phase of the beam relative to the cavity. In ER state, where the beam passes through the same cavity twice, this expression becomes:
\begin{eqnarray}
P_b^{\text{ER}} &=& P_+ - P_- - P_c = I_{\uparrow} \Delta V_{\uparrow} + I_{\downarrow} \Delta V_{\downarrow} \\
&\equiv& P_{b,\uparrow} + P_{b,\downarrow}\nonumber
\label{eq:imperfectER1}
\end{eqnarray}
$I_{\uparrow}$ and $\Delta V_{\uparrow}$, and $I_{\downarrow}$ and $\Delta V_{\downarrow}$ denote the current and voltage gain of the beam on the accelerating and decelerating pass through cavity respectively.  Note that this convention implies $\Delta V_{\downarrow}<0$. In the case of perfect ER $I_{\uparrow}=I_{\downarrow}$ (no beam loss), $\Delta V_{\downarrow} = \Delta V_{\uparrow}$ (no timing error), and the recovered power from the beam equals the power delivered during acceleration $P_{b,\uparrow}=-P_{b,\downarrow}$. In addition to minimizing beam loss, realizing this perfect ER condition requires careful setting of the cavity phases as well as tight control of the time of flight around the loop.  Determining the cavity phases must include the effects of the beam being non-relativistic. The power balance efficiency $\varepsilon_{P}$ per cavity, as well the power balance efficiency of the full 1-turn configuration can thus be defined as:
\begin{eqnarray}
\varepsilon_{P} &=& 1 - \frac{P_b^{\text{ER}}}{P_{b,\uparrow}}
\label{eq:eff_P1}
\\
\varepsilon_{P}^{\text{1-turn}} &=& 1 - \frac{\langle P_b^{\text{ER}} \rangle}{\langle P_{b,\uparrow} \rangle},
\label{eq:eff_P2}
\end{eqnarray}
where the average is taken by summing over all six MLC cavities.  

The power balance efficiencies described above provide a useful measure of the effectiveness of an established ER state from an operations perspective: when taken with the maximum power deliverable for each cavity, these efficiencies determine the maximum achievable current in assuming no beam loss. However, they do not directly quantify the efficiency of ER for a single particle that is not lost in the system.  In light of this, we define the per cavity and 1-turn single particle ER efficiency as:
\begin{eqnarray}
\varepsilon_{\text{ER}} &=& -\frac{\Delta V_{\downarrow}}{\Delta V_{\uparrow}},
\label{eq:eff_ER1}
\\
\varepsilon_{\text{ER}}^{\text{1-turn}} &=& -\frac{\langle \Delta V_{\downarrow}\rangle}{\langle \Delta V_{\uparrow}\rangle}.
\label{eq:eff_ER2}
\end{eqnarray}
The above ER efficiencies are related to the power balance efficiencies via:
\begin{eqnarray}
\varepsilon_{P} &=& \varepsilon_{\text{ER}}\left(1-
\frac{\Delta I}{I_{\uparrow}}\right),
\label{eq:eff_relations1}
\\
\varepsilon_{P}^{\text{1-turn}} &=& \varepsilon_{\text{ER}}^{\text{1-turn}}\left(1-
\frac{\Delta I}{I_{\uparrow}}\right),
\label{eq:eff_relations2}
\end{eqnarray}
where $\Delta I = I_{\uparrow}-I_{\downarrow}$.

In order to use the expressions in Eqs.~(\ref{eq:eff_P1}-\ref{eq:eff_relations2}), measurements of the various beam loading terms must be performed. In practice, the measurement of forward and reflected powers is subject to imperfect isolation of the dual directional coupler used to measure them. Accordingly, a forward travelling wave will excite a signal both in the forward coupled port and the reverse coupled port, while the reverse travelling wave will do the same. It can be shown in this case that, the difference between imperfectly measured forward and reflected power is given by,
\begin{equation}
    \tilde P_+-\tilde P_-=\tilde P_c + \chi\Delta V I + \mathrm{O}(I^2) \,,
    \label{eq:pdiff}
\end{equation}
where the tilde represent the fact that the quantities generally differ from those in equation (1) due to the imperfect nature of the dual directional coupler. Note that the term $\tilde P_c$ is a constant independent of beam current $I_b$ and $\chi$ is a constant independent of beam current and cavity voltage. The quadratic term here is negligible when we are in the regime of $\frac{I_b}{V_c}\frac{R}{Q}Q_L << 1$, where $Q_L$ is the loaded quality factor of the cavity \cite{ref:Banerjee2020}. Based on the cavity parameters used in CBETA \cite{ref:CDR}, we expect $\tilde P_+-\tilde P_-$ to be a linear function of beam current up to 10~$\mu\mathrm{A}$. The power delivered to the beam as determined by the dual directional coupler measurement is thus 
\begin{equation}
\tilde P_b = \chi \Delta V I + \mathrm{O}(I_b^2).
\end{equation}
\begin{figure*}[!t]
    \centering
    \includegraphics[scale=0.6]{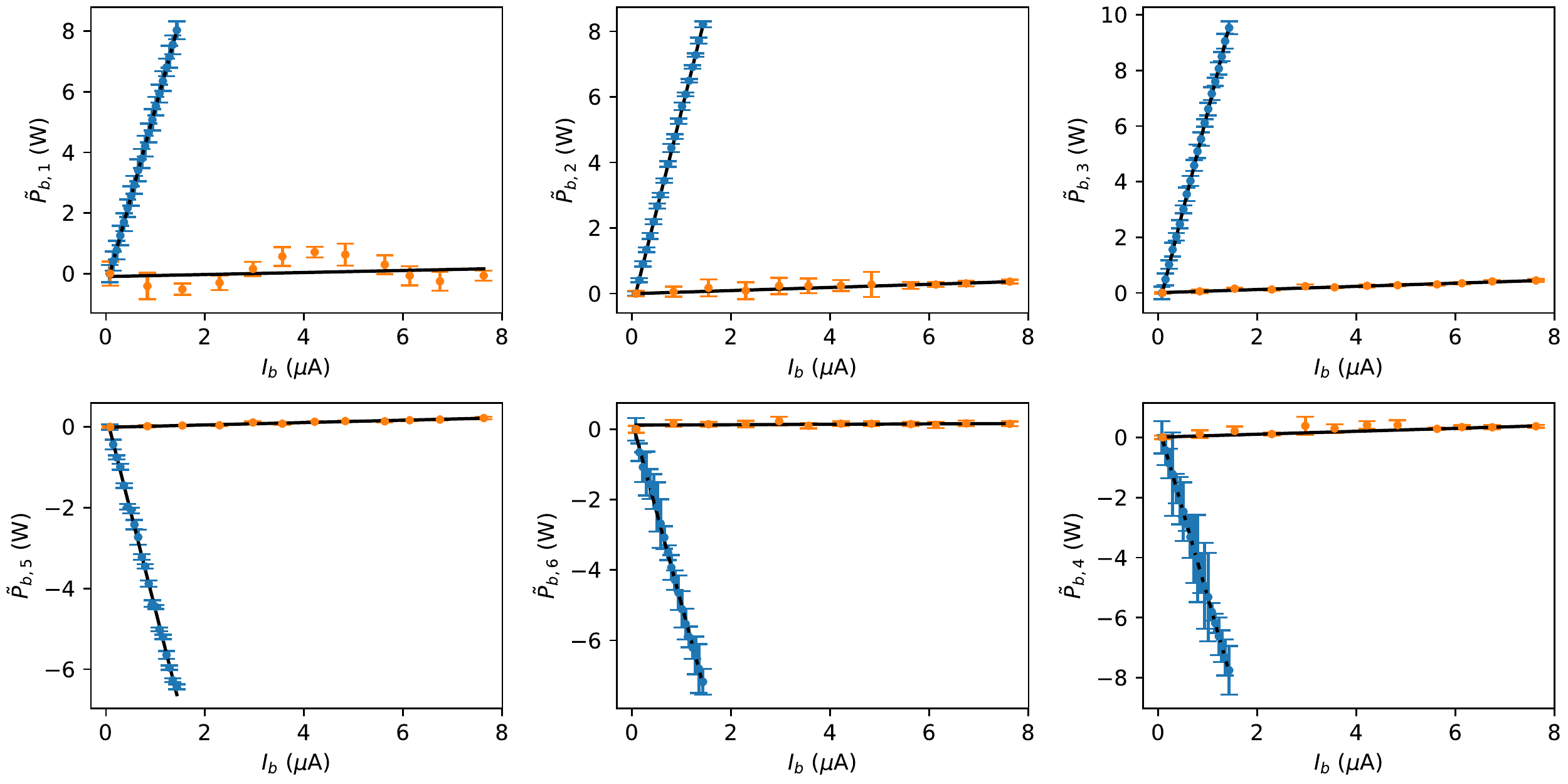}
    \caption{Beam loading as a function of current in all six cavities of the main linac. The blue data represent loading in the 3 up, 3 down configuration, clearly showing cavities the first three cavities transferring power to the beam while last three recover energy from the beam. The orange data show the beam loading during ER operation.  Linear fits to the data are shown in black and are used to extract the slope of the beam loading per unit current.}
    \label{fig:bl_data}
\end{figure*}
\begin{figure}[!htb]
\centering
\includegraphics*[width=\columnwidth]{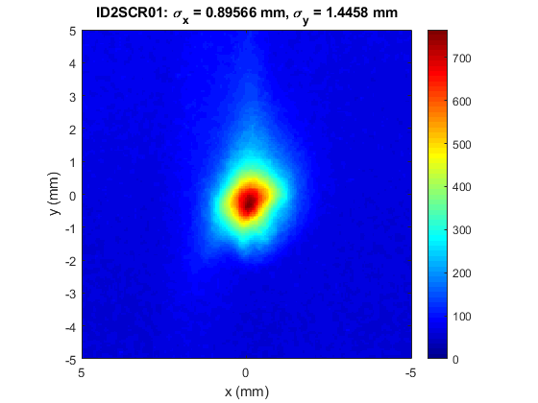}
\caption{Viewscreen of first recirculated beam on first screen in beam stop line.}
\label{FIG:FIRST_RECIRC}
\end{figure}

Figure~\ref{fig:bl_data} shows the beam loading data obtained during high-current operations of the 1-turn CBETA configuration. Two data sets were taken. The first is the $(3\uparrow,3\downarrow)$ configuration shown in blue in the plots.  Here the first three cavities of the MLC were configured to accelerate the beam by $\Delta V^\mathrm{(3\uparrow,3\downarrow)}\approx 6$ MV each and the next three decelerated the beam by the same amount. Consequently, the beam was transported from the injector through the MLC directly into the dump. The plots accordingly show a positive slope for the first three cavities transferring energy to the beam, while the other three show a negative slope indicating energy recovery. This measurement serves as a calibration measurement to estimate the constants $\chi_i = (\partial\tilde{P}_{b,i}/\partial I_b)/\Delta V$ as seen in Eq.~(\ref{eq:pdiff}).
%\begin{table*}[htb]
%\begin{center}
%\caption{Per Cavity Measurements}\label{tab:cparams}
%\begin{ruledtabular}
%\begin{tabular}{ c | c | c c c c c c}
% Cavity & 6 & 5 & 4 & 3 & 2 & 1\\
%\hline
%\hline
%$\chi$ & 0.99$\pm$0.02 & 1.004$\pm$0.001 & 1.18$\pm$0.01 & 0.96$\pm$0.05 & 0.809$\pm$0.001 & 0.87$\pm$0.03 \\
%\hline
%%$\epsilon_b$ & 99.2 & 99.2 & 99.2 & 99.2 & 99.4 & 99.9\\
%$\varepsilon_{P_b}$ & 99.2 & 99.2 & 99.2 & 99.2 & 99.4 & 99.9\\
%$\epsilon_{\text{ER}}$ & 99.9 & 99.9 & 99.8 & 99.8 & 100 & 101 \\
%\end{tabular}
%\end{ruledtabular}
%\end{center}
%\end{table*}
\begin{table*}[htb]
\begin{center}
\caption{Per Cavity Measurements and Predictions}\label{tab:cparams}
\begin{ruledtabular}
\begin{tabular}{ c | c c c c c c c}
 Cavity & 1 & 2 & 3 & 4 & 5 & 6\\
\hline
\hline
$\chi$ & $0.989\pm0.028$ & $1.005\pm0.006$ & $1.18\pm0.018$ & $0.924\pm0.047$ & $0.782\pm0.006$ & $0.85\pm0.033$ \\
\hline
$\varepsilon_{P_b}$ (\%) & $99.36\pm0.51$ & $99.28\pm0.19$ & $99.23\pm0.07$ & $99.18\pm 0.14$ & $99.40\pm0.05$ & $99.84\pm0.24$\\
$\varepsilon_{\text{ER}}$ (\%) & $99.99\pm0.42$ & $99.91\pm0.19$ & $99.86\pm0.12$ & $99.81\pm0.15$ & $100.04\pm0.11$ & $100.48\pm0.23$\\
\hline
Model $\varepsilon_{\text{ER}}$ (\%) & 100.20 & 100.07 & 100.01 & 99.98 & 99.94 & 99.72
\end{tabular}
\end{ruledtabular}
\end{center}
\end{table*}
Table~\ref{tab:cparams} shows the resulting dual directional coupler coefficients. The estimated error here reflects the error in the slopes, and does not include the error in the assumed energy gain/loss per cavity. 

Another important observation that comes from the fact that the beam loading in Figure~\ref{fig:bl_data} are linear with the beam current as measured in the DC gun.  Assuming minimal beam is lost in the injector, then $I_b=I_{\uparrow}$ and the power balance efficiencies can be written as:
\begin{eqnarray}
\varepsilon_{P} &=& 1 - \frac{\partial P_b^{\text{ER}}/\partial I}{\partial P_{b,\uparrow}/\partial I}
\label{eq:eff_P1_slopes}
\\
\varepsilon_{P}^{\text{1-turn}} &=& 1 - \frac{\langle \partial P_b^{\text{ER}/} \partial I\rangle}{\langle \partial P_{b,\uparrow}/\partial I \rangle}.
\label{eq:eff_P2_slopes}
\end{eqnarray}
Unfortunately, the quantity $P_{b,\uparrow}$ was not measured while the machine was in the ER state, so this term must be inferred from other data.  To do so we note that 
\begin{eqnarray}
\frac{\partial P_{b,\uparrow}}{\partial I}=\Delta V_{\uparrow}\approx V_c\cos(\phi)(1-\delta V),
\end{eqnarray}
where $\delta V$ is used to represent some small correction due to non-relativistic effects.  Neglecting this term and noting that the phase offset here is less than a few degrees, and thus $\cos\phi\approx1$ gives:
\begin{eqnarray}
\varepsilon_{P} &=& 1 - \frac{1}{V_c}\frac{\partial P_b^{\text{ER}}}{\partial I}
\label{eq:eff_P1_approx}
\\
\varepsilon_{P}^{\text{1-turn}} &=& 1 - \frac{1}{\langle V_c\rangle}\frac{ \partial \langle P_b^{\text{ER}}\rangle }{\partial I }.
\label{eq:eff_P2_approx}
\end{eqnarray}
Note that the factor $\chi$ cancels out of the first of these two equations, but is needed to evaluate the second expression above.  To evaluate the ER efficiency terms in Eqs.~(\ref{eq:eff_ER1}, \ref{eq:eff_ER2}) requires knowledge of $\Delta V_{\downarrow}$ which is not directly measured.  Instead, we note that the dipole magnet for sending the decelerated beam to the beam stop was set for 6 MeV for these measurements, indicating $\varepsilon_{\text{ER}}^{\text{1-turn}}\approx1$.  %(0.9997 as predicted by the MLC model). 
Fig.~\ref{FIG:FIRST_RECIRC} shows an image of the energy recovered beam at 6 MeV in the dump line, demonstrating $\varepsilon_{\text{ER}}^{\text{1-turn}}\approx1$. Using this and substituting Eq.~(\ref{eq:eff_relations2}) into Eq.~(\ref{eq:eff_relations1}) gives 
\begin{eqnarray}
\varepsilon_{\text{ER}} = \varepsilon_P\left(\frac{\varepsilon_{\text{ER}}^{\text{1-turn}}}{\varepsilon_P^{\text{1-turn}}}\right)\approx\varepsilon_P/\varepsilon_P^{\text{1-turn}},
\end{eqnarray}
which can be computed using the results from Eqs.~(\ref{eq:eff_P1_approx}--\ref{eq:eff_P2_approx}).  The second to last two rows of Table~\ref{tab:cparams} show the values for these quantities for each cavity.  The range of the per cavity power balance efficiency is 99.2--99.8\%, while the range of the per cavity ER efficiency is 99.8--100.5\%.  The latter shoes the correct trend when compared to the model values shown in the last row of Table~\ref{tab:cparams}.

\begin{figure}[!htb]
\centering
\includegraphics*[width=0.75\columnwidth]{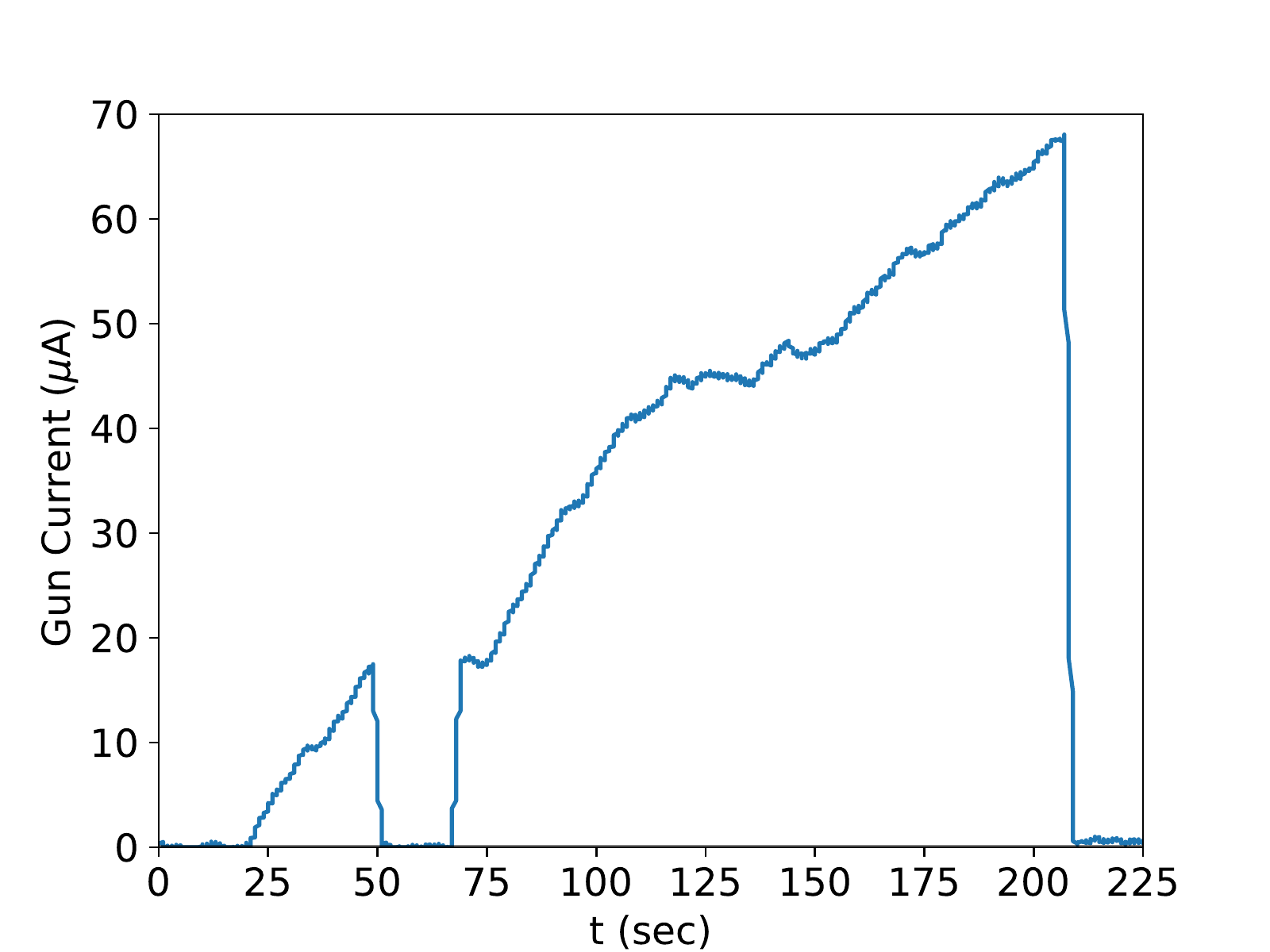}
\caption{Plot of average beam current in gun during higher current push.}
\label{fig:current}
\end{figure}
After performing the cavity loading measurements the current in the machine was increased. During most of the commissioning period the average beam current in CBETA was restricted to a few nA using 5 pC bunches.  After producing nearly 10 $\mu$A during ER measurements, the beam current was steadily increased to see what, if any limitations in current could be found.  Fig.~\ref{fig:current} shows the results of this current scan. The maximum current achieved was roughly 70 $\mu$A using 2.5 pC bunches.  The radiation pattern on both the internal and external monitors at the time suggested that currents up to 1~mA were achievable, however at the time it was decided not to increase the current any further as the fast shutdown system designed to protect the machine in high mode was not fully installed yet.

\section{\label{sec4:level1}Conclusions}

The single turn CBETA configuration was successfully set up for nearly perfectly balanced energy recovery. Direct measurements of the beam loading in the main linac cavities demonstrate a high single particle energy recover efficiency of all six SRF linac cavities. In particular a total one-turn power balance efficiency of 99.4\% was measured.  The corresponding per cavity power balances and single particle energy recovery efficiencies ranged from 99.2--99.8\% and 99.8--100.5\%, respectively.  These values roughly agree with predicted values from particle tracking in the first five out of six MLC cavities.

\begin{acknowledgments}
This work was supported by NSF Grant No. DMR-0807731, DOE Award No. DE-SC0012704, and NYSERDA Contract No. 102192.
\end{acknowledgments}

\bibliography{bibliography}% Produces the bibliography via BibTeX.

\end{document}